\documentclass[epj]{svjour}
\usepackage{latexsym}
\usepackage{graphics}
\usepackage{graphicx}
\usepackage{mathtools}
\usepackage{slashed}
\usepackage{esvect}
\usepackage{amsmath}
\usepackage{mathrsfs}
\usepackage{newtxtext}
\usepackage{physics}
\usepackage{xcolor}
\begin{document}
	\title{Path integral formalism for the free Dirac propagator in spherical coordinates}
	\author{Sreya Banerjee \and Zolt\'an Harman
	}
	\institute{Max Planck Institute of Nuclear Physics, Saupfercheckweg 1, 69117 Heidelberg, Germany}
	\date{\today}
	\abstract{
	The relativistic Green's function of a free spin-$\frac{1}{2}$ fermion is derived using the Feynman path integral formalism in spherical coordinates.
	The Green's function is reduced to an exactly solvable path integral by an appropriate coordinate transformation.
	The result is given in terms of spherical Bessel functions and spherical spinors, and agrees with previous solutions of the problem.
	}
	
	\maketitle
	
	\section{Introduction}
	\label{intro}
	
	Green's functions of the Dirac equation has been used in different areas of atomic physics: in the calculation of radiative corrections in atoms and highly charged
	ions~\cite{Mohr1998,Wichmann1956,Gyulassy1975,Snyderman1991,Pachucki1993,Indelicato1998,Yerokhin1999,Moskovkin2004,Yerokhin2020,Yerokhin2021},
	tested by various state-of-the-art experimental methods (see e.g.~\cite{Gumberidze2005,Gillaspy2010,Sturm2011,Kubicek2014,Beyer2017,Indelicato2019});
	in describing relativistic atomic processes~\cite{Humphries1984,Koval2003}, multiphoton interactions~\cite{Maquet1998}, x-ray scattering~\cite{Surzhykov2015,Wong1985},
	atoms in external fields~\cite{Szmytkowski1997}, and the weak decay in muonic atoms~\cite{Szafron2016}, to name a few examples.
	
	In the current article, we derive the free Dirac Green's function using the path integral formalism of Feynman.
	The Green's function is reduced in Biedenharn's basis~\cite{Biedenharn1962} into a radial path integral, the effective action of which is similar to that of
	a non-relativistic particle. In order to express the energy-dependent Green's function in a closed form, we convert the radial path integral to the path integral of an isotropic harmonic oscillator through coordinate transformation
	along with local time rescaling, in analogy with earlier works of Inomata and collaborators on the hydrogen atom~\cite{Inomata1984,Kayed1984}.
	The final results agree with the solution of the inhomogeneous Dirac equation in previous works, e.g. in Refs.~\cite{Mohr1998,Mohr1974}.
	
	\section{First- and second-order Dirac equation}
	\label{sec:1}

	The time-independent Dirac equation is customarily expressed, in natural units
	($c=1$, $\hbar=1$), as
	\begin{equation}
		(E-\boldsymbol{\alpha} \cdot \hat{\boldsymbol{p}}-\beta m) \Psi=0\,.
		\label{eq:1}
	\end{equation}
	Here, $E$ is the energy, $m$ is the mass of a particle, and $\hat{\boldsymbol{p}}$ is the operator of 3-momentum.
	The $\boldsymbol{\alpha} = (\alpha_1, \alpha_2, \alpha_3)$ and $\beta$ are the usual $4\times 4$ Dirac matrices, and $\Psi$ is the bispinor
	wave function.
	The Green's function $G(\boldsymbol{r}_2,\boldsymbol{r}_1; E)$ depends on the two positions $\boldsymbol{r}_1$, $\boldsymbol{r}_2$ and the energy, and it
	satisfies the inhomogeneous Dirac equation
\begin{equation}
	(m-\hat{M})G(\boldsymbol{r}_2,\boldsymbol{r}_1; E)=\delta(\boldsymbol{r}_2 - \boldsymbol{r}_1) \,,
	\label{eq:2}
\end{equation}
	where $\hat{M}=-\beta\boldsymbol{\alpha} \cdot \hat{\boldsymbol{p}}+\beta E$.
	This Green's function can be expressed as~\cite{Manakov_1997,Kayed1984}
\begin{equation}
	G(\boldsymbol{r}_2,\boldsymbol{r}_1; E)=(m+\hat{M})g(\boldsymbol{r}_2, \boldsymbol{r}_1; E) \,,
	\label{eq:3}
\end{equation}
	where $g(\boldsymbol{r}_2,\boldsymbol{r}_1; E)$ is the solution of the second-order -- or iterated -- inhomogeneous Dirac equation
\begin{equation}
	(m^2-\hat{M}^2)g(\boldsymbol{r}_2,\boldsymbol{r}_1; E)=\delta(\boldsymbol{r}_2 - \boldsymbol{r}_1)\,.
	\label{eq:4}
\end{equation}
	The second-order Dirac equation resembles the Schr\"odinger equation, and its solution has a simpler form than the usual first-order
	equation~\cite{Biedenharn1962}.

	We use spherical coordinates to find explicit expressions for these Green's functions. Generally, operators of interest can be
	rewritten using the Dirac operator $\hat{K}=\beta(\hat{\boldsymbol{\Sigma}} \cdot \hat{\boldsymbol{L}}+1)$,
	the radial momentum operator $\hat{p}_r=\frac{1}{r}(\boldsymbol{r} \cdot \hat{\boldsymbol{p}}-i)$, and
	${\alpha}_r=\frac{\boldsymbol{\alpha} \cdot \boldsymbol{r}}{r}$, which is the component of the Dirac matrix $\boldsymbol{\alpha}$
	in the direction $\boldsymbol{r}$. Here, the operator $\hat{\boldsymbol{\Sigma}}=\mathbf{1}\otimes\boldsymbol{\sigma}$, with 
	$\mathbf{1}$ being the $2 \times 2$ unit matrix and
	$\boldsymbol{\sigma}$ the vector of the Pauli matrices, defines the Dirac spin operator $\hat{\boldsymbol{S}}_{\rm D}=\frac{1}{2}\hat{\boldsymbol{\Sigma}}$.
	As usual, $\hat{\boldsymbol{L}}$ is the operator of orbital angular momentum, and the eigenvalue of $\hat{\boldsymbol{L}}^2$ is $l(l+1)$, with $l$ being
	an integer.
	
	The operator $\hat{K}$ commutes with the first-order Dirac Hamiltonian and with $\alpha_r$. The Dirac operator is related to the total angular momentum
	operator~\cite{Bjorken1964} $\hat{\boldsymbol{J}}$ via $\hat{K}^2=\hat{\boldsymbol{J}}^2+\frac{1}{4}$. The operator $\hat{\boldsymbol{J}}^2$ has the
	eigenvalues $j(j+1)$, with $j$ being half-integer, and thus the eigenvalues of $\hat{K}^2$ are $(j+\frac{1}{2})^2$. The eigenvalues of $\hat{K}$ are
\begin{equation}
	\kappa=\mp\left(j + \frac{1}{2}\right)\,,
	\label{eq:5}
\end{equation}
	for $j=l \pm \frac{1}{2}$.

\section{Path integral form of the Green's function}

	We seek the solution to equation (\ref{eq:5}) using the integral representation
\begin{equation}
	g(\boldsymbol{r}_2,\boldsymbol{r}_1; E) \equiv \bra{\boldsymbol{r}_2}\hat{g}\ket{\boldsymbol{r}_1}=
	\frac{i}{2m}\int_{0}^{\infty}\bra{\boldsymbol{r}_2}e^{i \hat{H} u}\ket{\boldsymbol{r}_1}\, du \,.
	\label{eq:6}
\end{equation}
	In the above equation, the integration is done with respect to the time-like parameter $u$ and the integrand
	$\bra{\boldsymbol{r}_2}e^{i {H} u}\ket{\boldsymbol{r}_1}$  can be interpreted as a propagator that describes a system evolving with the
	parameter $u$ from $\boldsymbol{r}_1$ to $\boldsymbol{r}_2$, and has an effective Hamiltonian, $\hat{H}$. As stated by Feynman,
	propagators can be written in the form of path integrals~\cite{Feynman2010}. Here, we express the integral in Eq.~(\ref{eq:6}) in terms
	of a path integral. Finally, we evaluate the Green's function for the Dirac equation by using Eq.~(\ref{eq:4}).
	
	The effective Hamiltonian can be written as
\begin{equation}
	\hat{H}=\frac{1}{2m}(m^2-\hat{M}^2)\,.
	\label{eq:8}
\end{equation}
	With the operators introduced above, following Biedenharn~\cite{10.1007/BFb0012286}, $\hat{M}$ can be written as
\begin{equation}
	\hat{M}=-\beta(\alpha_r \hat{p}_r)+i\frac{\alpha_r \hat{K}}{r}+\beta E \,,
	\label{eq:9}
\end{equation}
	and the effective Hamiltonian (\ref{eq:8}) can be cast in the form
\begin{equation}
	\hat{H}=\frac{1}{2m}\left(\hat{p}_r^{2}+\frac{\hat{K}(\hat{K}-\beta)}{r^2}-E^2+m^2\right)\,.
	\label{eq:10}
\end{equation}
	In Eq.~(\ref{eq:10}), the coefficient of the $1/r^2$ term in the Hamiltonian is analogous to the term containing the orbital angular momentum term in the
	Hamiltonian of the Schr\"odinger equation. To establish this correspondence, we use a specific case of the Martin-Glauber operator~\cite{Martin1958}
\begin{equation}
	\hat{\mathscr{L}}=-\beta \hat{K},
\end{equation}
	for which the following holds: $\hat{\mathscr{L}}^2=\hat{K}^2$. The eigenvalue of $\hat{\mathscr{L}}$, $\gamma$, is also given as
\begin{eqnarray}
	\gamma         = \mp \left(j+\frac{1}{2}\right) \,.
	\label{eq:11}
\end{eqnarray} 
	We define $\lambda$ as a function of $\gamma$, $\lambda \equiv |\gamma|+\frac{1}{2}(\mbox{sign}\gamma-1)$, such that the eigenvalue equation~\cite{Martin1958}
	\begin{equation}
		\hat{\mathscr{L}} \left(\hat{\mathscr{L}} +1 \right) \ket{\lambda}=\lambda(\lambda+1)\ket{\lambda} \,.
		\label{eq:15}
	\end{equation}
	is fulfilled. The angular wave functions $\braket{\theta\phi}{\lambda} = \braket{\theta\phi}{j,\mu,\kappa,\tilde{\beta}}$ are the simultaneous eigenstates of the operators
	$\hat{\mathscr{L}}(\hat{\mathscr{L}} +1 )$, $\hat{\boldsymbol{J}}^2$, $\hat{J}_z$, $\hat{K}$ and the $\beta$ matrix,
	with the eigenvalues $\lambda(\lambda+1)$, $j(j+1)$, $\mu$, $\kappa$ and $\tilde{\beta}=\pm 1$, respectively.
	They can be written in the explicit bispinor form as~\cite{Biedenharn1962,Kayed1984}
	\begin{eqnarray}
		&\braket{\theta\phi}{j,\mu,\kappa,\tilde{\beta}=-1}=
		\begin{pmatrix}
			0\\
			\chi_{\kappa}^{\mu}
		\end{pmatrix},\nonumber\\
		&\braket{\theta\phi}{j,\mu,\kappa,\tilde{\beta}=1}=
		\begin{pmatrix}
			\chi_{-\kappa}^{\mu}\\
			0
		\end{pmatrix}\,,\label{eq:19}
	\end{eqnarray}
	where the $\chi_{\kappa}^{\mu}$ are the well-known spherical spinors~\cite{Bjorken1964}
	\begin{equation}
		\chi_{\kappa}^{\mu}=\sum_{\mu' = \pm \frac{1}{2}} C\left(l\,,\frac{1}{2}\,,j; \mu-\mu'\,, \mu'\,, \mu \right) Y_l^{\mu-\mu'}(\theta,\phi)
		\chi_{\frac{1}{2}}^{\mu'}\,,
		\label{eq:20}
	\end{equation}
	expressed in terms of the Clebsch-Gordan coefficients $C(\dots)$, the spherical harmonics $Y_l^{\mu-\mu'}$, and the unit spinors
	\begin{eqnarray}
		\chi_{\frac{1}{2}}^{\frac{1}{2}} =
		\begin{pmatrix}
			1\\
			0
		\end{pmatrix}\,, \quad
		\chi_{\frac{1}{2}}^{-\frac{1}{2}} =
		\begin{pmatrix}
			0\\
			1
		\end{pmatrix}\,.
	\end{eqnarray}
	
	The Hamiltonian of Eq.~(\ref{eq:10}) is now represented in this basis~as
\begin{equation}
	\hat{H}_{\lambda}=\frac{\hat{p}_r^2}{2m}+\frac{\lambda(\lambda+1)}{2m r^2}-\frac{k^2}{2m}\,,
	\label{eq:16}
\end{equation}
	where $k^2=E^2-m^2$.

	Having defined the radial Hamiltonian, the Green's function for the second-order Dirac equation can be expressed as a partial wave expansion
	\begin{equation}
		\bra{\boldsymbol{r}_2}\hat{g}\ket{\boldsymbol{r}_1}=\sum_{\lambda}\bra{\theta_2 \phi_2}\ket{\lambda}\bra{r_2}
		\hat{g}_{\lambda}\ket{r_1}\bra{\lambda}\ket{\theta_1 \phi_1}\,.
		\label{eq:17}
	\end{equation}
	Here, the radial Green's function is
	\begin{equation}
		\bra{{r}_2} \hat{g}_{\lambda}\ket{{r}_1}=\frac{ i }{2m}\int\,\bra{{r}_2}e^{- i  \hat{H}_{\lambda}u}\ket{{r}_1}du \,,
		\label{eq:18}
	\end{equation}
	where the operator $\hat{H}_{\lambda}$ is defined in Eq.~ (\ref{eq:16}).
	This radial Hamiltonian undergoes a evolution in $u$, and is very similar in form to the radial propagator
	of the non-relativistic hydrogen atom as, introduced by Inomata~\cite{Inomata1984}.
	
	The Green's function given in Eq.~(\ref{eq:17}) can now be reduced using Eq. ~(\ref{eq:16}-\ref{eq:18}), yielding
	\begin{equation}
		\bra{\boldsymbol{r}_2} \hat{g}\ket{\boldsymbol{r}_1}=\sum_{j,\kappa}\bra{{r}_2} \hat{g}_{\lambda}\ket{{r}_1}
		\Omega_{\kappa,\kappa^{'}}^j(\theta_2\phi_2|\theta_1\phi_1)\beta^2 \,,
		\label{eq:19}
	\end{equation}
	where
	\begin{equation}
		\Omega_{\kappa,\kappa^{'}}^j(\theta_2\phi_2|\theta_1\phi_1)=\sum_{\mu}\chi_{\kappa}^{\mu}(\theta_2,\phi_2)\chi_{\kappa^{'}}^{\mu\dagger}(\theta_1,\phi_1)\,.
		\label{eq:20}
	\end{equation}

	In order to proceed with the construction of the path integral in spherical coordinates, we establish the partial action term for a very short time interval
	$t_j=u_j-u_{j-1}$. This partial action can be approximated as $S(\vb{r}_j,\vb{r}_{j-1})\approx t_jL(\Delta \vb{r}_j/t_j,\vb{r}_{j})$, where
	$\Delta \vb{r}_j=\vb{r}_j-\vb{r}_{j-1}$.The total path is divided into $N$ intervals
	by local time rescaling such that $r_0=r_1$, $r_N=r_2$, and $u=\sum t_j$. The radial action term for the Hamiltonian in Eq.~(\ref{eq:16}) is given as
		\begin{eqnarray}
		S(t_j)=\frac{m(\Delta r_j)^2}{2t_j}-\frac{\lambda(\lambda+1)t_j}{2mr_jr_{j-1}}+\frac{(E^2-m^2)t_j}{2m}\,.
		\label{eq:21}
	\end{eqnarray}
Since angular motion has rotational symmetry, we concern ourselves with only the radial motion associated with this action. As such, the corresponding radial function is
\begin{eqnarray}
 &&	R_\lambda(r_j,r_{j-1})=\frac{i t_j}{2mr_jr_{j-1}}\nonumber\\
 &&\times \exp\left\{\frac{im(\Delta r_j)^2}{2t_j}-\frac{it_j\lambda(\lambda+1)}{2mr_jr_{j-1}}+\frac{ik^2t_j}{2m}\right\}\,.
 \label{eq:22}
\end{eqnarray}
Summing over the Feynman histories, the radial propagator for the radial function in Eq.~(\ref{eq:22}), is obtained as
\begin{eqnarray}
	\label{eq:23}
	&&K_\lambda(r_2,r_1;u)=\\
	&&\lim_{N\rightarrow\infty}\int \prod_{j=1}^{N} \{R_\lambda(r_j, r_{j-1})\}\prod_{j=1}^{N}\left[\frac{m}{2\pi i t_j}\right]^\frac{3}{2}\prod_{j=1}^{N-1}(r^2 dr)\,. \nonumber
\end{eqnarray}
  This expression for the radial propagator can be represented as~\cite{Inomata1984,Kayed1984}
 \begin{eqnarray}
 	\label{eq:24}
 	&&K_\lambda(r_2,r_1;u)=\bra{{r}_2}e^{- i  \hat{H}_{\lambda}u}\ket{{r}_1}=(r_1 r_2)^{-1}\\
 	&&\times \lim_{N\rightarrow\infty}
 	\int\,\exp\left[ i \sum_{j=1}^N S(t_j)\right]\prod_{j=1}^N\left[\frac{m}{2\pi i  t_j}\right]^{\frac{1}{2}}\prod_{j=1}^{N-1}dr_j\,. \nonumber
 \end{eqnarray}
In order to solve the path integral in Eq.~(\ref{eq:23}), we simplify the radial function in Eq.~(\ref{eq:22}).
Taking all contributions in $t_j$ upto first order in consideration, for small $t_j$, we can apply the approximation formula 
 \begin{eqnarray}
 	\label{eq:25}
 	&&I_\nu\left(\frac{n}{t_j}\right)=\\
 	&&\left(\frac{2\pi n}{t_j}\right)^{-\frac{1}{2}}\exp\left\{\frac{n}{t_j}-\frac{1}{2}\left[\left(\nu^2-\frac{1}{4}\right)\frac{t_j}{n}+\mathscr{O}(t_j^2)\right]\right\}\nonumber\,,
\end{eqnarray}
which reduces the radial function to the form
\begin{eqnarray}
	\label{eq:26}
	&&R_\lambda(r_j,r_{j-1})=\left(\frac{i\pi t_j}{2mr_jr_{j-1}}\right)^{\frac{1}{2}}\\
	&&\times\exp\left[\frac{im(r_j^2+r^2_{j-1})}{2t_j}+\frac{ik^2t_j}{2m}\right]I_{\lambda+\frac{1}{2}}\left(\frac{mr_jr_{j-1}}{it_j}\right)\,.\nonumber
\end{eqnarray} 
Subsituting this expression in Eq.~(\ref{eq:23}), we obtain
\begin{eqnarray}
	\label{eq:27}
	&&K_\lambda(r_2,r_1;u)=(r_1r_2)^{-\frac{1}{2}}\left(\frac{-im}{u}\right)\\
	&&\times\exp\left\{\frac{ik^2u}{2m}\right\}\exp\left\{\frac{1}{2}\frac{im(r_1^2+r_2^2)}{u}\right\}I_{\lambda+\frac{1}{2}}\left(\frac{mr_jr_{j-1}}{iu}\right)\nonumber\,.
\end{eqnarray}
Eq.~(\ref{eq:27}) gives the radial propagator for the radial Hamiltonian and when substituted in Eq.~(\ref{eq:18}) yields the radial Green's function in the form
\begin{eqnarray}
	\label{eq:28}
	&&\bra{{r}_2}g_{\lambda}\ket{{r}_1}=(r_1r_2)^{-\frac{1}{2}}\int\left(\frac{-im}{u}\right)\\
	&&\times\exp\left\{\frac{ik^2u}{2m}\right\}\exp\left\{\frac{1}{2}\frac{im(r_1^2+r_2^2)}{u}\right\}I_{\lambda+\frac{1}{2}}\left(\frac{mr_jr_{j-1}}{iu}\right)\,du\nonumber\,.
\end{eqnarray}
However, the integration on the right hand side of Eq.~(\ref{eq:28}) does not have a known closed-form solution. Thus, to enable the calculation process, and to bring the integral in Eq.~(\ref{eq:28}) to a reducible form, following~\cite{Inomata1984}, we modify the radial action in Eq.~(\ref{eq:21}), such that it represents the action of a three-dimensional isotropic harmonic oscillator, by replacing the radial variable $r_j$ by $\rho_j=\sqrt{r_j}$ and the local time-slicing parameter $t_j$ by $\sigma_j=t_j/4\bar{r}_j$,
where the geometric mean is $\bar{r}_j=\sqrt{r_j r_{j-1}}=\rho_j\rho_{j-1}=\bar{\rho}_j^2$. For small values of $t_j$, the geometric mean, $\bar{r}_j$, gives the mid-point value which is well-defined for a classical path. 
	The action term can now be represented as
	\begin{eqnarray}
		S(\tau_j)&=&\frac{m(\Delta \rho_j)^2}{2\sigma_j}+\frac{m(\Delta \rho_j)^4}{8\sigma_j\bar{\rho_j}^2}-\frac{2\lambda(\lambda+1)\sigma_j}{m\bar{\rho}_j^2} \\
		&& -\frac{1}{2}m\omega^2\bar{\rho_j}^2\sigma_j\,,\nonumber
		\label{eq:29}
	\end{eqnarray}
	with $\omega=\frac{2 i  k}{m}$. The measure of the integrand in the Eq.~(\ref{eq:24}) also changes according to the transformed variables,
	it is expressed as
	\begin{eqnarray}
		\prod_{j=1}^N\left[\frac{m}{2\pi i  t_j}\right]^{\frac{1}{2}}\prod_{j=1}^{N-1}dr_j=
		\frac{1}{\sqrt{4\rho_1\rho_2}}\prod_{j=1}^N\left[\frac{m}{2\pi i \sigma_j}\right]^{\frac{1}{2}}\prod_{j=1}^{N-1}d\rho_j \,. \nonumber
	\end{eqnarray}
	Despite these modifications, we encounter a different problem that makes the integration of Eq.~(\ref{eq:24}) difficult; the second term in the modified radial
	action expression contains $\rho_j$ raised to its fourth power which causes the integral to diverge.
	Rewriting Eq.~(\ref{eq:23}) after making the necessary substitutions, we obtain
	\begin{eqnarray}
		&&\bra{{r}_2}e^{- i  \hat{H}_{\lambda}u}\ket{{r}_1}= \\
		&&(\rho_1\rho_2)^{-2}\lim_{N\rightarrow\infty}\int\exp\Biggl[ i \sum_{j=1}^N\frac{m(\Delta\rho_j)^2}{2\sigma_j}+\frac{m(\Delta\rho_j)^4}{8\sigma_j(\bar{\rho}_j)^2}\nonumber\\
		&&-\frac{2\lambda(\lambda+1)\sigma_j}{m(\bar{\rho}_j)^2}
		-\frac{1}{2}m(\omega)^2(\bar{\rho}_j)^2\sigma_j \Biggr] \nonumber \\
		&&\times (4\rho_1\rho_2)^{-\frac{1}{2}}\prod_{j=1}^N\left[\frac{m}{2\pi i \sigma_j}\right]^{\frac{1}{2}}\prod_{j=1}^{N-1}d\rho_j \,. \nonumber
		\label{eq:30}
	\end{eqnarray}
	In order to overcome the problem posed by $\rho_j^4$, we use the integral formula~\cite{Inomata1984} that is valid for large $A$ and integer $n$:
	\begin{eqnarray}
		&&\int x^{2n}\exp[-Ax^2+Bx^4+\mathscr{O}(x^6)]dx=\\
		&&\int x^{2n}\exp[-Ax^2+\frac{3}{4} BA^{-2}+\mathscr{O}(A^{-3})]dx \,. \nonumber
		\label{eq:31}
	\end{eqnarray}
	This allows the fourth-order term to be represented by a replacement term given by $-3\sigma_j/(8m\bar{\rho}_j^2)$, yielding
	\begin{eqnarray}
		&&\bra{{r}_2}e^{- i  \hat{H}_{\lambda}u}\ket{{r}_1}= \\
		&&(\rho_1\rho_2)^{-2}\lim_{N\rightarrow\infty}\int\exp\Biggl[ i \sum_{j=1}^N\frac{m(\Delta\rho_j)^2}{2\sigma_j}+\frac{3\sigma_j}{8m(\bar{\rho}_j)^2}\nonumber\\
		&&-\frac{2\lambda(\lambda+1)\sigma_j}{m(\bar{\rho}_j)^2}
		- \frac{1}{2}m(\omega)^2(\bar{\rho}_j)^2\sigma_j \Biggr] \nonumber \\
		&&\times(4\rho_1\rho_2)^{-\frac{1}{2}}\prod_{j=1}^N\left[\frac{m}{2\pi i \sigma_j}\right]^{\frac{1}{2}}\prod_{j=1}^{N-1}d\rho_j \,. \nonumber
		\label{eq:32}
	\end{eqnarray}
	We express this equation in a compact form as
	\begin{eqnarray}
		\label{eq:33}
		\bra{{r}_2}e^{- i \hat{H}_{\lambda}u}\ket{{r}_1}=\frac{1}{2}(\rho_1\rho_2)^{-\frac{3}{2}}\tilde{K}_\lambda(\rho_2,\rho_1;\sigma) \,, 
	\end{eqnarray}
	where the propagator $\tilde{K}$ of the $\sigma$ evolution is defined as
	\begin{eqnarray}
		\label{eq:34}
		&&\tilde{K}_\lambda(\rho_2,\rho_1;\sigma)=(\rho_1\rho_2)^{-1} \\
		&&\times \lim_{N\rightarrow\infty}\int\exp\left[ i \sum_{j=1}^N\tilde{S}(\sigma_j)\right]
		\prod_{j=1}^N\left[\frac{m}{2\pi i \sigma_j}\right]^{\frac{1}{2}}\prod_{j=1}^{N-1}d\rho_j \,,  \nonumber
	\end{eqnarray}
	and the modified action term is now given as
	\begin{equation}
		\tilde{S}(\sigma_j)=\frac{m(\Delta\rho_j)^2}{2\sigma_j}-\frac{\lambda^{'}(\lambda^{'}+1)\sigma_j}{2m(\bar{\rho}_j)^2}-\frac{1}{2}m\omega^2(\bar{\rho}_j)^2 \sigma_j \,,
		\label{eq:35}
	\end{equation}
	where $\lambda^{'}=2\lambda+\frac{1}{2}$.
	
	This effective action term is analogous to the radial action term of a three-dimensional harmonic oscillator. Thus the propagator in Eq.~(\ref{eq:18}),
	evolving with $u$, has also been reduced to the propagator of a harmonic oscillator, and can be evaluated by following the procedure introduced by
	Inomata and Peak~\cite{Peak1969}. Thus we obtain the radial propagator, in terms of the modified Bessel function $I_{\nu}(x)$ with $\nu=\lambda^{'}+\frac{1}{2}$, as
	\begin{eqnarray}
		\label{eq:36}
		&&\tilde{K}_\lambda(\rho_2,\rho_1;\sigma) =- i (\rho_1\rho_2)^{-\frac{1}{2}}(m\omega)\csc(\omega\sigma) \\
		&&\exp\left[\frac{1}{2} i  m\omega\left((\rho_1^2+\rho_2^2\right)\cot(\omega\sigma)\right]
		I_{\lambda^{'}+\frac{1}{2}} \left(\frac{m}{i}\omega\rho_1\rho_2\csc(\omega\sigma) \right)\,. \nonumber
	\end{eqnarray}
It is to be noted that in the limit that $\omega$ reduces to zero, the propagator in Eq.~(\ref{eq:36}) reduces to that defined in Eq.~(\ref{eq:27}),
which is the propagator for a free particle in three dimensions. However, since we are concerned with determining the energy-dependent Green's function
in a closed form, we proceed with a finite-valued $\omega$.
	
	Substituting the radial propagator from the above equation into Eq.~(\ref{eq:33}) yields
	\begin{eqnarray}
		\label{eq:37}
		&&\bra{{r}_2}e^{- i \hat{H}_{\lambda}u}\ket{{r}_1}=\frac{1}{2}(r_1 r_2)^{-1}(2k) \\
		&&\times \csc\left(\frac{ i  kt}{m (r_1 r_2)^{\frac{1}{2}}}\right) \exp\left[-k(r_1+r_2)\cot\left(\frac{ i  kt}{m(r_1 r_2)^{\frac{1}{2}}}\right)\right] \nonumber \\
		&&\times I_{2\lambda+1}\left( 2k(r_1 r_2)^{\frac{1}{2}}\csc\left(\frac{ i  kt}{m(r_1 r_2)^{\frac{1}{2}}}\right)\right)\,. \nonumber 
	\end{eqnarray}
	Using this result for the integrand in Eq.~(\ref{eq:18}) we obtain
	\begin{eqnarray}
		\label{eq:38}
		\bra{r_2} \hat{g}_\lambda \ket{r_1}&=&(r_1r_2)^{-\frac{1}{2}}\int\exp[ i k(r_1+r_2)\coth q] \\
		&\times& I_{2\lambda+1}\left(-2 i  k(r_1r_2)^{\frac{1}{2}}\csch q \right)\csch q\,dq \,,\nonumber
	\end{eqnarray}
	where
	\begin{eqnarray}
	q = \frac{kt}{(4r_1r_2m^2)^{\frac{1}{2}}}\,.\nonumber
	\end{eqnarray}
	The type of integral in Eq.~(\ref{eq:38}) has a closed solution~\cite{Gradshtein1981}:
	\begin{eqnarray}
		\label{eq:39}
		&& \int \exp[ i   k(r_1+r_2)\coth q] \\
		&& \times I_{2\lambda+1} \left( -2 i  k(r_1r_2)^{\frac{1}{2}}\csch q \right)\csch q\,dq=\nonumber\\
		&& \frac{\Gamma(\lambda+1)}{2 i  k(r_1r_2)^{\frac{1}{2}}\Gamma(2\lambda+2)} \nonumber \\
		&& \times M_{0,\lambda+\frac{1}{2}}(-2 i  kr_2)W_{0,\lambda+\frac{1}{2}}(-2 i  kr_1)\,, \nonumber
	\end{eqnarray}
	where $\Gamma$ is the gamma function, and $M$ and $W$ are the Whittaker functions.
	They can be expressed in terms of the modified Bessel functions~\cite{Abramowitz1964}:
	\begin{eqnarray}
		M_{0,\nu}(2z)&=&2^{2\nu +\frac{1}{2}} \Gamma(1+\nu)\sqrt{z}I_{\nu}(z) \,,\nonumber\\
		W_{0,\nu}(2z)&=&\sqrt{\frac{2z}{\pi}}K_\nu(z)\,. \nonumber
	\end{eqnarray}
	Substituting these into Eq.~(\ref{eq:39}), we obtain
	\begin{eqnarray}
		\label{eq:40}
		&&\bra{r_2} \hat{g}_\lambda\ket{r_1}=\frac{\Gamma(\lambda+1)}{2 i  kr_1r_2\Gamma(2\lambda+2)} 2^{2\lambda+\frac{3}{2}} \\
		&&\times \Gamma \left(\lambda+\frac{3}{2}\right)\sqrt{-i kr_2}I_{\lambda+\frac{1}{2}}(-i kr_2) \nonumber\\
		 &&\times \sqrt{\frac{-2ikr_1}{\pi}}K_{\lambda+\frac{1}{2}}(-ikr_1)\,. \nonumber
	\end{eqnarray}
	Using this expression for the radial part in the expression for the total Green's function of the iterated Dirac equation yields
\begin{eqnarray}
	\label{eq:41}
	&&\bra{\boldsymbol{r}_2} \hat{g} \ket{\boldsymbol{r}_1}=
	\sum_{j,\kappa}\frac{\Gamma(\lambda+1)}{2 i  kr_1r_2\Gamma(2\lambda+2)} 2^{2\lambda+\frac{3}{2}} \\
	&&\times \Gamma \left(\lambda+\frac{3}{2}\right)\sqrt{-i kr_2}I_{\lambda+\frac{1}{2}}(-i kr_2) \nonumber \\
	&&\times \sqrt{\frac{-2ikr_1}{\pi}}K_{\lambda+\frac{1}{2}}(-ikr_1) \Omega_{\kappa,\kappa^{'}}^j(\theta_2\phi_2|\theta_1\phi_1)\beta^2\,. \nonumber
	\end{eqnarray}
	The modified Bessel functions of the first and second kind can be expressed in terms of spherical functions~\cite{Abramowitz1964}, and so we obtain
\begin{eqnarray}
	\label{eq:42}
	&&\bra{\boldsymbol{r}_2} \hat{g} \ket{\boldsymbol{r}_1}=
	\sum_{j,\kappa}\frac{\Gamma(\lambda+1)}{2 i  kr_1r_2\Gamma(2\lambda+2)} 2^{2\lambda+\frac{3}{2}}\\
	&&\times \Gamma\left(\lambda+\frac{3}{2}\right)\sqrt{-i kr_2}(i)^{-\left(\lambda+\frac{1}{2}\right)}\sqrt{\frac{2ikr_2}{\pi}}j_{\lambda}(ikr_2) \nonumber \\
	&&\times \sqrt{\frac{-2ikr_1}{\pi}}\frac{\pi}{2}(i)^{\lambda+\frac{3}{2}}\sqrt{\frac{2ikr_1}{\pi}}h_{\lambda}(ikr_1)\Omega_{\kappa,\kappa^{'}}^j(\theta_2\phi_2|\theta_1\phi_1)\beta^2\,,\nonumber
\end{eqnarray}
	where $j_{\lambda}$ and $h_{\lambda}$ are the spherical Bessel and Hankel functions, respectively.

	The operator $\hat{M}$, when acting on a state with a given $\kappa$, takes the form
\begin{equation}
	i\beta\alpha_r\left[\frac{\partial}{\partial r}+\frac{1-\gamma\beta}{r}\right]+\frac{\kappa E}{\gamma}\beta\,.
	\label{eq:43}
\end{equation}
	We can turn to calculating the free Dirac Green's function from Eq.~(\ref{eq:4}):
\begin{eqnarray}
	\label{eq:44}
	&&\bra{\boldsymbol{r}_2} \hat{G} \ket{\boldsymbol{r}_1}=
	\sum_{j,\kappa}\frac{\Gamma(\lambda+1)}{2 i  kr_1r_2\Gamma(2\lambda+2)} 2^{2\lambda+\frac{3}{2}}\nonumber\\
	&&\times \Gamma\left(\lambda+\frac{3}{2}\right)\sqrt{-i kr_2}(i)^{-\left(\lambda+\frac{1}{2}\right)}\sqrt{\frac{2ikr_2}{\pi}}\nonumber\\
	&&\times\sqrt{\frac{-2ikr_1}{\pi}}\frac{\pi}{2}(i)^{\lambda+\frac{3}{2}}\sqrt{\frac{2ikr_1}{\pi}}h_{\lambda}(ikr_1)\nonumber\\
	&&\times \left(m+i\beta\alpha_r\left[\frac{\partial}{\partial r}+\frac{1-\gamma\beta}{r}\right]+\frac{\kappa E}{\gamma}\beta\right)j_{\lambda}(ikr_2)\nonumber\\
	&&\times \Omega_{\kappa,\kappa^{'}}^j(\theta_2\phi_2|\theta_1\phi_1)\beta^2\,.
\end{eqnarray}
	Using the relation for the derivative of spherical Bessel functions~\cite{Abramowitz1964} one obtains
\begin{eqnarray}
	\label{eq:45}
	&&\bra{\boldsymbol{r}_2} \hat{G} \ket{\boldsymbol{r}_1}=
	\sum_{j,\kappa}k(2\lambda+2)i h_{\lambda}(ikr_1)\\
	&&\times\biggl\{\left(m+\frac{\kappa E}{\gamma}\beta\right)j_{\lambda}(ikr_2) \Omega_{\kappa,\kappa^{'}}^j(\theta_2\phi_2|\theta_1\phi_1)\beta^2\nonumber\\
	&&+ i\left[k\left\{\frac{\lambda}{ikr_2}j_{\lambda}(ikr_2)-j_{\lambda+1}(ikr_2)\right\}+\frac{(1\pm\gamma)}{r_2} j_{\lambda}(ikr_2)\right]\nonumber\\
	&&\times\Omega_{\kappa,\kappa^{'}}^j(\theta_2\phi_2|\theta_1\phi_1)\beta^{2}\alpha_r\biggr\}\,.\nonumber
\end{eqnarray}
	On account of the relations
\begin{eqnarray}
	\alpha_r   &=& i\sigma_r\beta\gamma^{1}\gamma^{2}\gamma^{3}\,,\nonumber\\
	\gamma^{i} &=& \beta\alpha_i\,, \quad \mbox{with } i \in \left\{ 1,2,3 \right\}\,, \nonumber \\
	\sigma_r\Omega^{j}_{\kappa,\kappa'} &=& -\Omega^{j}_{\kappa,-\kappa'}\,. \nonumber
\end{eqnarray}
	the Green's function of the first-order free Dirac equation can be finally written as
\begin{eqnarray}
	\label{eq:46}
	&&\bra{\boldsymbol{r}_2} \hat{G} \ket{\boldsymbol{r}_1}=
	\sum_{j,\kappa}ik(2\lambda+2) h_{\lambda}(ikr_1) \nonumber \\
	&&\times \biggl\{\left(m+\frac{\kappa E}{\gamma}\beta\right)j_{\lambda}(ikr_2) \Omega_{\kappa,\kappa^{'}}^j(\theta_2\phi_2|\theta_1\phi_1)\beta^2\nonumber\\
	&& -\tilde{\beta}\left[k\left(\frac{\lambda}{ikr_2}j_{\lambda}(ikr_2)-j_{\lambda+1}(ikr_2)\right)+\frac{(1\pm\gamma)}{r_2} j_{\lambda}(ikr_2)\right]\nonumber\\
	&&\times\Omega_{\kappa,-\kappa}^j(\theta_2\phi_2|\theta_1\phi_1)\alpha_1\alpha_2\alpha_3\biggr\}\,.
\end{eqnarray}
	This formula is equivalent to the free Green's function derived by different methods~\cite{Mohr1998,Yerokhin1999,Mohr1974}.

	\section{Summary}
	\label{sec:summary}
	
	The free Dirac Green's function has been derived from first principles within the path integral formalism. Spherical coordinates were used, in order
	to arrive to a form applicable in atomic physics calculations. The Green's function has been transformed in Biedenharn's basis~\cite{Biedenharn1962}
	into a radial path integral, with an effective action resembling the action of the Schr\"odinger equation.
	The radial path integral has been converted through coordinate transformation and with a time rescaling to that of a classical isotropic harmonic
	oscillator which reduces the problem to an exactly solvable form. The final result is expressed with spherical Bessel functions and spherical spinors
	in Eq.~(\ref{eq:46}).
	
	\section*{Acknowledgements}
	
	This work is funded by the Deutsche Forschungsgemeinschaft (DFG, German Research Foundation) -- Project-ID 273811115 -- SFB 1225 (ISOQUANT).
	
	\section*{Author contributions}
	
	S.~B. and Z.~H. conceived the model, interpreted the results and wrote the paper. S.~B. performed most of the calculations, in consultation with Z.~H.
	All authors gave final approval for publication.
	
	\subsection*{Data Availability Statement} This manuscript has no associated data or the data will not be deposited.

\end{document}